\documentstyle[a4,12pt]{article}

\begin{document}
\sloppy

\newcommand{\zb}{\bar z}
\newcommand{\th}{\theta}
\newcommand{\tb}{\bar{\theta}}
\newcommand{\zt}{\tilde z}
\newcommand{\zbt}{\tilde{\zb}}
\newcommand{\tht}{\tilde{\th}}
\newcommand{\tbt}{\tilde{\tb}}
\newcommand{\TH}{\Theta}
\newcommand{\TB}{\bar{\Theta}}

\newcommand{\la}{\lambda}
\newcommand{\LA}{\Lambda}
\newcommand{\DE}{\Delta}
\newcommand{\OM}{\Omega}
\newcommand{\NA}{\nabla}

\newcommand{\CC}{{\cal C}}
\newcommand{\CD}{{\cal D}}
\newcommand{\CW}{{\cal W}}
\newcommand{\CR}{{\cal R}}
\newcommand{\CS}{{\cal S}}

\newcommand{\pa}{\partial}
\newcommand{\pab}{\bar{\partial}}
\newcommand{\dab}{\bar D}

\newcommand{\pat}{\tilde{\pa}}
\newcommand{\pabt}{\tilde{\pab}}
\newcommand{\dt}{\tilde{D}}
\newcommand{\dbt}{\tilde{\dab}}
\newcommand{\de}{\delta}
\newcommand{\db}{\bar{\de}}

\newcommand{\bz}{{\bf z}}
\newcommand{\bZ}{{\bf Z}}

\newcommand{\ZB}{\bar Z}
\newcommand{\HT}{{H_{\th}}^z}
\newcommand{\HB}{{H_{\tb}}^z}
\newcommand{\HO}{H_{\th}^{\ \th}}
\newcommand{\HZ}{H_{\zb}^{\ \th}}
\newcommand{\HZB}{H_{\zb}^{\ z}}
\newcommand{\HOB}{H_{\tb}^{\ \th}}

\newcommand{\ds}{\displaystyle}

\newcommand{\TT}{{\cal T}_{\th z}}
\newcommand{\TZ}{{\cal T}_{\tb \zb}}

\newtheorem{theo}{Theorem}[section]
\newtheorem{lem}{Lemma}[section]
\newtheorem{prop}{Proposition}[section]
\newtheorem{cor}{Corollary}[section]

%%%%%%%%%%%%%%%%%%%%%%% begin of text %%%%%%%%%%%%%%%%%%%%%%%%%%%

\hfill MPI-Ph/92-66
\vskip 0.07truecm
\hfill KA-THEP-7/92

\thispagestyle{empty}

\bigskip
\bigskip
\begin{center}
{\bf \Huge{Superconformally Covariant Operators}}
\end{center}
\begin{center}
{\bf \Huge{and}}
\end{center}
\begin{center}
{\bf \Huge{Super $W$ Algebras}}
\end{center}
\bigskip
\bigskip
\centerline{{\bf Fran\c cois Gieres$^{\, a \, \S}$} $\ $  and $\ $
{\bf Stefan Theisen$^{\, b \, \ddag}$}}
\bigskip
\bigskip
\centerline{$^{a}${\it Max-Planck-Institut f\"ur Physik}}
\centerline{\it Werner-Heisenberg-Institut}
\centerline{\it F\"ohringer Ring 6}
\centerline{\it D - 8000 - M\"unchen 40}
\bigskip
\bigskip
\centerline{$^{b}${\it Institut f\"ur Theoretische Physik}}
\centerline{\it Universit\"at Karlsruhe}
\centerline{\it Kaiserstra\ss e 12}
\centerline{\it D - 7500 - Karlsruhe 1}
\bigskip
\bigskip
\bigskip
\bigskip
\bigskip

\begin{abstract}

We study superdifferential
operators of order $2n+1$ which are covariant with respect
to superconformal changes of coordinates on a compact super
Riemann surface.
We show that all such operators arise from super M\"obius covariant ones.
A canonical matrix representation is presented and
applications to classical super $W$ algebras are discussed.

\end{abstract}
\bigskip
\bigskip
\bigskip

\nopagebreak
\begin{flushleft}
\rule{2in}{0.03cm} \\

{\footnotesize \ ${}^{\S}$
Alexander von Humboldt Fellow.
E-mail address: frg@dm0mpi11.}
\\  [-0.04cm]
{\footnotesize \ ${}^{\ddag}$
E-mail address: be01@dkauni2.}
\\  [0.5cm]

MPI-Ph/92-66   \\
KA-THEP-7/92    \\
\vskip 0.07truecm
August 1992
\end{flushleft}
\newpage
\setcounter{page}{1}

\section{Introduction}

The study of linear $n$-th order differential operators in one complex
variable which are conformally covariant represents a classic subject
in the mathematical literature \cite{bol} \cite{gp}. In recent years,
these topics have regained
considerable interest and a variety of applications
in mathematics and physics have been discussed (see \cite{cco}
for a partial review and further references).
In particular, it was realized
\cite{pan} \cite{dfiz} \cite{bx}
that these operators
give rise to classical $W$ algebras.

The natural supersymmetric extension of this subject consists
of the study of operators of the form $D^{2n+1} +...$ (where
$D \equiv \pa / \pa \th \, + \, \th \, \pa / \pa z$
and $n = 0,1, 2,...$)
which are defined on compact super Riemann surfaces.
The subclass of
these operators which only depends on the projective structure
(and not on additional variables)
has been investigated in detail in reference
\cite{cco}: these are the so-called
super Bol operators. In the present work, we will be
concerned with the most general operators of order $2n+1$
which are superconformally covariant. Along the lines of reference \cite{dfiz},
we will study their general structure,
their classification and discuss
the applications to classical super $W$ algebras.

This paper is organized as follows. After introducing the necessary tools
and notation in section 2,
we discuss the general form of covariant
operators of order $2n+1$. In the sequel, specific subclasses of these
operators are constructed by starting from operators which are
covariant with respect to superprojective changes of coordinates. The first
examples are provided by the super Bol operators ${\cal L}_n$
mentioned above and the second
class is given by operators $M_{W_k} ^{(n)}$
(with $1 \leq k \leq 2n+1$)
which do not only depend on the superprojective
structure, but also on some superconformal fields $W_k$.
(In the applications
to $W$ algebras, the projective structure is related to the super stress
tensor while the conformal fields $W_k$
correspond to the currents for the
$W$ symmetries.)
By adding the operators
$M_{W_k} ^{(n)}$ to the super Bol operator
${\cal L}_n$, we obtain again a covariant
operator of order $2n+1$ :
\begin{equation}
\label{gen}
{\cal L}
\ = \ {\cal L}_n \, + \, M_{W_1} ^{(n)}  \, + \, ... \, + \,
M_{W_{2n+1}} ^{(n)}
\ \ \ .
\end{equation}
In section six we will show that this already encompasses
the most general case. In other words,
all operators of order $2n+1$
which are superconformally covariant can be cast into the form (\ref{gen}).
Thereafter, it is shown that the operators $M_{W_k}^{(n)}$
represent special cases of operators which are bilinear and covariant
and which are of independent interest. However, in
the sequel, we return to the linear operators and we present a
matrix representation for them, thereby elucidating the underlying algebraic
structure. Section 9 is devoted to the application of the
previous results to the description of classical super $W$ algebras.
In fact, the aforementioned matrix representation provides a convenient
set of generators for these algebras and the Poisson brackets
between these generators involve the previously constructed covariant
operators. While our main discussion is concerned with superfields,
the derivation of component field results is addressed in section 10.
We conclude with some remarks on topics which are closely
related to our subject (covariant operators
of even degree, in $N=2$ supersymmetry and
in higher dimensions,
singular vectors of the Neveu-Schwarz algebra).
In an appendix we collect some of the algebraic concepts which are
referred to in the main text.

\section{General framework}

Let us first recall the notions which are needed in the sequel \cite{df}.
The arena we work on, is a compact
$N=1$ super Riemann surface (SRS) parametrized by
local coordinates $\bz=(z,\theta)$.
(Our discussion applies equally well to a real one-dimensional supermanifold
for which case the
changes of coordinates are superdiffeomorphisms.)
The canonical derivatives are denoted by
$\partial=\partial_z$ and
$D=\partial_\theta+ \theta\partial$.
By definition, any two sets of local coordinates on the SRS,
$\bz$ and
$\bz^{\prime}$, are related by a superconformal transformation
$\bz \to \bz^\prime(\bz)$, i.e. a transformation satisfying
$D z^{\prime} = \theta^{\prime} (D \th^\prime )$.
This condition implies
$D = (D \th^\prime) D^{\prime}$
and
$(D \th^\prime)(D^\prime \th ) =1$.

Throughout the text, the Jacobian of the superconformal change of coordinates
$\bz \to \bz^\prime(\bz)$ will be denoted by
\[
{\rm e}^{-w} \ \equiv \ D \th^{\prime}
\ \ .
\]

By ${\cal F}_n$ we denote
the space of superconformal fields
of weight ${n\over 2}$ on the SRS, i.e.
superfields
with transformation properties
$C_n(\bz)\to C_n^{\, \prime} (\bz^\prime)
= {\rm e}^{nw} \, C_n(\bz)$.
The field $C_n$ is taken to have Grassmann parity $(-)^n$.

The super Schwarzian derivative of the coordinate transformation
$\bz \to \bz^\prime(\bz)$
is defined by
\begin{equation}
\label{53}
{\cal S}(\bz^\prime , \bz )
\ = \ - [  D^3 w + ( \pa w  ) ( Dw )  ]
\ = \
{\pa^2 \th^{\prime} \over D \th^\prime}
- 2 \, {(\pa \th^{\prime})
(D^3 \th^{\prime}) \over (D \th^\prime )^2}
\ \ .
\end{equation}
Under the composition of superconformal transformations,
$\bz \to \bz^\prime \to \bz^{\prime \prime}$, it transforms according to
\begin{equation}
\label{54}
{\cal S}(\bz^{\prime \prime},\bz ) \ = \
{\rm e}^{-3w} \,
{\cal S}(\bz^{\prime \prime} ,\bz^{\prime} ) + {\cal S}(\bz^{\prime} , \bz)
\ \ ,
\end{equation}
which relation implies
${\cal S}(\bz^{\prime} ,\bz) \ = \
- {\rm e}^{-3w} \, {\cal S}(\bz,\bz^{\prime} )$.

Coordinates belonging to a superprojective atlas on the SRS
will be denoted by capital
letters, $ \bZ = ( Z , \TH )$. They are related to each other
by superprojective (super M\"obius) transformations, i.e.
superconformal changes of coordinates
$\bZ \to \bZ^{\prime} (\bZ)$
for which ${\cal S}(\bZ^{\prime} ,\bZ) = 0$.
Direct integration of this equation and of
$D_{\TH} Z^{\prime}  = \TH^\prime ( D_{\TH} \TH^{\prime})$ gives
\begin{eqnarray}
\label{47}
Z^\prime  & = & \frac{aZ+b}{cZ+d}
\, + \, \TH\, {\gamma Z+\delta\over(cZ+d)^2}
\qquad\quad(ad-bc=1)
\\
\TH^\prime  & = & {\gamma Z+\delta\over cZ+d}
\, + \, \TH \, {1+{1\over 2}\delta\gamma\over cZ+d}
\ \ .
\nonumber
\end{eqnarray}
Here, $a,b,c,d$ are even and $\gamma, \de$ odd constants;
we redefined the parameters in such a way that the
even part of the transformation for $Z$ coincides with
ordinary projective transformations.
The associated Jacobian then reads
\[
D_{\TH} \TH^{\prime} \ = \
( \tilde{c} Z + \tilde{d} + \TH \tilde{\gamma} )^{-1}
\]
with
$\tilde{c} = c \, ( 1 - \frac{1}{2} \,  \delta \gamma)
\, , \,
\tilde{d} = d \, ( 1 - \frac{1}{2} \, \delta \gamma)
\, , \,
\tilde{\gamma} = c\delta - d \gamma$.

\section{The most general covariant operators}

The most general superdifferential operator which is linear, superanalytic
and of order $2n+1$ has the local form
\[
{\cal L} \ = \ a_0 \, D^{2n+1} + a_1 \, D^{2n}
+ a_2 \, D^{2n-1} + ... + a_{2n+1}
\ \ ,
\]
where the coefficients $a_p \equiv a_p^{(n)} (\bz)$
are analytic superfields. We will always take $a_p$ to have
Grassmann parity $(-)^p$.
If the leading coefficient $a_0$ does not have any zeros, one can achieve
$a_0 \equiv 1$ by dividing by this coefficient.
In the sequel, we will make this choice and
we will study this type of operator on the SRS
from the point of view of
superconformal changes of coordinates.

The requirement that ${\cal L}$ maps superconformal fields
(of a generic weight ${p\over 2}$)
again to superconformal fields, i.e.
\[
{\cal L}  \ : \ {\cal F}_{p}
\longrightarrow \ {\cal F}_{p+2n+1}
\ \ ,
\]
determines the transformation laws of the coefficients $a_1 ,..., a_{2n+1}$
under a
superconformal change of coordinates $\bz\to\bz^\prime(\bz)$. For the first two
coefficients, one finds that
\begin{eqnarray}
\label{52a}
a_1 ^\prime
& = &  {\rm e}^{w} \,
\biggl\{ \,  a_1 - (p+n) \,  ( D w ) \, \biggr\}
\\
a_2 ^\prime
& = &  {\rm e}^{2w} \,
\biggl\{ \, \biggl[ a_2 - n(p+n) (\pa w)\biggr]
+ n \,  ( D w ) \,  a_1 \, \biggr\}
\ \ .
\nonumber
\end{eqnarray}
{}From the first of these equations,
we see that $a_1$ is a superconformal field
of weight ${1\over 2}$ if $p =-n$.
In that case, $a_1$ can be eliminated from ${\cal L}$ by
performing the rescaling
\begin{equation}
\label{50a}
{\cal L} \ \longrightarrow \ g^{-1} {\cal L} \, g
\quad \quad \quad
{\rm with}
\quad \quad
g( \bz ) \, = \, \exp \left\{ \, - \int_{\bz_0}^{\bz}
d\tilde{\bz} \, a_1 (\tilde{\bz} ) \, \right\}
\ \ ,
\end{equation}
where the integral is to be understood as an indefinite integral
\cite{df}, i.e. $D\int_{{\bf z}_0}^{\bf z} d\tilde{\bf z} \,
a_1 (\tilde{\bf z}) = a_1 ({\bf z})$.
As a result,
one obtains an operator of the form
\begin{equation}
\label{50}
{\cal L}^{(n)}=D^{2n+1}\,+\,\sum_{p=2}^{2n+1}a_p   \,D^{2n+1-p}
\ \ .
\end{equation}
{}From now on, we will consider this operator which
will be referred to as ``{\em normalized operator}$\,$" of order $2n+1$.

\begin{lem}
There are transformation laws of the coefficients $a_p$
under superconformal changes of coordinates $\bz\to\bz^\prime(\bz)$
such that
the operator ${\cal L}^{(n)}$ maps superconformal fields to
superconformal fields; more precisely,
\begin{equation}
{\cal L}^{(n)} \ : \ {\cal F}_{-n}
\longrightarrow \ {\cal F}_{n+1}
\ \ ,
\end{equation}
which entails that ${\cal L}^{(n)}$ transforms as
\begin{equation}
\label{51}
{\cal L}^{(n)}\,\to\,{\cal L}^{(n)\prime} = {\rm e}^{(n+1)w}
\, {\cal L}^{(n)} \,
{\rm e}^{nw}
\ \ .
\end{equation}
The transformation laws of the first few coefficients are explicitly given by
\begin{eqnarray}
a_2^\prime
&=&
{\rm e}^{2w} \,  a_2
\nonumber
\\
a_3^\prime
&=&
{\rm e}^{3w} \, \biggl\{ \,
\biggl[ a_3 - {n+1 \choose 2}   \, {\cal S} \biggr]
+  ( D w ) \,  a_2 \, \biggr\}
\label{52}
\\
a_4^\prime
& = &
{\rm e}^{4w}  \,
\biggl\{ \, \biggl[ a_4  - {n+1 \choose 3}  \, D {\cal S} \biggr]
+(n-1) (Dw) \biggl[ a_3  -  {n+1 \choose 2}  \, \CS \biggr]
(n-1) ( \pa w ) \, a_2 \, \biggr\}
\nonumber  \\
a_5^\prime
& = &
e^{5w} \, \biggl\{
\biggl[ a_5 - 2 {n+1 \choose 3} \pa \CS \biggr]
                 + 2 (Dw) \biggl[ a_4 - {n+1\choose 3} D \CS \biggr]
                 + 2 (n-1) (\pa w) \biggl[ a_3 - {n+1\choose 2} \CS \biggr]
\nonumber \\
           &   & \qquad
+ \frac{1}{2} (n-1) \biggl[ (n+2)(D^3 w)+(n+6)(\partial w)
                 (Dw) \biggr] a_2 \, \biggr\}
\nonumber \\
a_6^\prime
& = &
e^{6w} \, \biggl\{ \biggl[ a_6 - 2 {n+1\choose 4} D^3\CS \biggr   ]
        + (n-2) (Dw) \biggl[ a_5 - 2 {n+1\choose 3} \pa \CS \biggr]
\nonumber \\
           &   & \quad
               + 2 (n-2) \pa w \biggl[ a_4 - {n+1\choose 3} D \CS \biggr]
              + {n-1\choose 2} \biggl[ (D^3 )
                 + 5 (Dw) (\pa w) \biggr] \biggl[ a_3 -{n+1\choose 2} \CS
                 \biggr]
\nonumber \\
 &   & \quad \ \ + {1\over 6} (n-1)(n-2) \biggl[ (3-2n) (Dw)(D^3 w )
                 + (n+3) (\pa^2 w) + (n+9) (\pa w)^2 \biggr] a_2
               \, \biggr\}
\ ,
\nonumber
\end{eqnarray}
where
$a_p^{\prime} = a_p^{\prime} (\bz ^{\prime}) \, , \,
a_p = a_p (\bz)$ and where
${\cal S} \equiv
{\cal S} (\bz^{\prime} , \bz )$ denotes
the super Schwarzian derivative.
\end{lem}

Since the coefficient $a_2$ of ${\cal L}^{(n)}$
transforms homogeneously, it is possible to
set it to zero in a consistent way.
If we do so, we have
$a_3 ( \bz ) = \frac{1}{2}  \, n (n+1) \, {\cal R}_{z\th} (\bz )$, where
${\cal R}$ represents a superprojective (or
super Schwarzian) connection on the SRS \cite{cco}:
locally, the latter
is given by a collection of odd superfields ${\cal R}_{z\th}$
which are locally superanalytic
and which transform under a superconformal change
of coordinates according to
\begin{equation}
\label{48}
{\cal R}_{z^{\prime} \th^{\prime}} ( \bz^{\prime} ) \ = \  {\rm e}^{3w}
\left[ \, {\cal R}_{z\th} ( \bz )
- {\cal S} (\bz^{\prime} , \bz ) \, \right]
\ \ .
\end{equation}
In the general case ($a_2$ not identically zero),
we conclude from eqs.(\ref{52}) and (\ref{48}) that
\begin{equation}
\label{49a}
\tilde{a}_3 \ \equiv \ a_3 \, - {1\over 2} \, Da_2
\ = \
\frac{1}{2}  \, n (n+1) \, {\cal R}
\ \ .
\end{equation}

On a compact SRS, there is a one-to-one correspondence between superprojective
connections and superprojective structures (i.e. superprojective atlases).
This relation is expressed by \cite{cco}
\begin{equation}
\label{49}
{\cal R}_{z\th} ( \bz ) \ = \ {\cal S} (\bZ , \bz )
\ \ ,
\end{equation}
where $\bZ$ belongs to a projective coordinate system
and $\bz$ to a generic one.
Note that the quantity (\ref{49}) transforms as in eq.(\ref{48}) with
respect to a conformal change of $\bz$ and that it is inert under
a super M\"obius transformation of $\bZ$.

Since
$\tilde{a}_3 (\bz ) \propto {\cal S} (\bZ , \bz )$ (or
$a_3 (\bz ) \propto {\cal S} (\bZ , \bz )$ for $a_2 \equiv 0$),
this coefficient vanishes
if $\bz$ is chosen to belong to the same superprojective atlas
as $\bZ$.
In the next section, we will start from such an atlas and
define simple operators which are covariant with respect to super M\"obius
transformations. Then, we go over to a generic coordinate system and
recover conformally covariant operators.

\section{Ex.1: Super Bol operators}

In this section, we recall some results from reference \cite{cco}.
We start from a superprojective atlas with coordinates $\bZ, \bZ^{\prime}$
related by eqs.(\ref{47}).
A field ${\cal C}_n (\bZ )$
transforming covariantly with respect to these changes of coordinates,
${\cal C}_n^{\prime} (\bZ^{\prime} ) =
(D_{\TH} \TH^{\prime})^{-n} \,
{\cal C}_n (\bZ )$,
is called a quasi-primary field of weight $\frac{n}{2}$.

Obviously, the simplest normalized operator of order $2n+1$ is given
by $D_{\TH} ^{2n+1}$. For this operator,
we have the following result which can be proven by induction:
\begin{lem}[Super Bol Lemma]
For quasi-primary superfields ${\cal C}_{-n}$ of weight $-{n \over 2}$,
the field $D_\Theta^{2n+1}{\cal C}_{-n}$ is quasi-primary
of weight ${n+1 \over 2}$, i.e.
it transforms under projective changes of coordinates according to
\begin{equation}
(D_\Theta^{2n+1}{\cal C}_{-n})^\prime=(D_\Theta\Theta^\prime)^{-(n+1)}
\; D_\Theta^{2n+1}{\cal C}_{-n}
\ \ .
\end{equation}
\end{lem}

We now go over from the projective coordinates
$\bZ$ to generic coordinates
$\bz$ by a superconformal transformation $\bz \to \bZ (\bz )$. Then,
$D_\TH^{2n+1}$ becomes the so-called
super Bol operator ${\cal L}_n$ acting on the conformal field
$C_{-n}(\bz ) $:
\begin{equation}
\label{44}
D_\Theta^{2n+1}{\cal C}_{-n} \ = \ (D\TH )^{-(n+1)} \,  {\cal L}_n
C_{-n}
\quad \quad \quad
{\rm with}
\quad \quad
{\cal C}_{-n} (\bZ )\ \equiv \ (D\TH )^n \,  C_{-n} (\bz )
\ \ .
\end{equation}
In operatorial form, this relation reads
\begin{eqnarray}
{\cal L}_n&=&(D\Theta)^{n+1}\left({1\over D\Theta}D\right)^{2n+1}
(D\Theta)^n\nonumber\\
&=&[D-nB][D-(n-1)B]\cdots[D+nB]
\ \ ,
\label{45}
\end{eqnarray}
where we introduced the quantity $B = D\, \ln \, D\Theta$.

\begin{cor}
The super Bol operator
${\cal L}_n$ as defined by eq.(\ref{44}) represents a normalized and
conformally covariant operator of order $2n+1$. It
depends only on the superprojective structure, i.e. it depends
on $B$ only through the
superprojective connection
\begin{equation}
\label{197}
{\cal R}(\bz) = {\cal S}(\bZ,\bz)=\partial B-BDB
\ \ .
\end{equation}
The explicit expression for ${\cal L}_n$ has the form
\begin{eqnarray}
{\cal L}_n \ = \ D^{2n+1}
& + &
{1\over 2}n(n+1){\cal R}D^{2n-2}
                 +{1\over 6}n(n^2-1)(D{\cal R})D^{2n-3}
                 \nonumber \\
           & + &  {1\over 3}n(n^2-1)(\partial{\cal R})D^{2n-4}
                 +{1\over 12}n(n^2-1)(n-2)(D^3 {\cal R})D^{2n-5}
                 \nonumber \\
           & + &  {1\over 8}n(n^2-1)(n-2)\left[\left(
                 \pa^2{\cal R}\right)+{1\over 3}(2n+3){\cal R}
                 (D{\cal R})\right]D^{2n-6}\,+\,\dots
\end{eqnarray}
\end{cor}

\noindent
For later reference, we display the first few super Bol
operators:
\begin{eqnarray}
{\cal L}_0 & = &  D
\nonumber  \\
{\cal L}_1 & = &
D^3  \, + \,  {\cal R}
\nonumber \\
{\cal L}_2 & = &
D^5   \, + \, 3\, {\cal R} \, \pa  \, + \,
(D {\cal R}) \, D \, + \,
2 \, (\pa {\cal R} )
\label{898}
 \\
{\cal L}_3 & = &
D^7   \, + \, 6\, {\cal R} \, \pa ^2  \, + \,
4\, (D {\cal R}) \, D^3 \, + \,
8 \, (\pa {\cal R} ) \, \pa \, + \,
2 \, ( D^3 {\cal R} ) \, D \, + \,
3 \, [\pa^2 {\cal R}  + 3 {\cal R} D {\cal R} ]
\nonumber
\\
{\cal L}_4 & = &
D^9   \, + \, 10 \, {\cal R} \, \pa ^3  \, + \,
10 \, (D {\cal R}) \, D^5 \, + \, 20 \, (\pa {\cal R} ) \, \pa^2
\, + \,
10 ( D^3 {\cal R}) D^3  \, + \, 5  [3 \pa^2 {\cal R}
 +
11 {\cal R} D {\cal R} ] \pa
\nonumber
\\
& & \, + \,
[ 3 \, D^5  {\cal R}  \, + \, 9 \, (D {\cal R} )^2 \, + \,
(\pa {\cal R}) {\cal R} ] \,  D
\, + \,
4\, [ \pa^3 {\cal R}  \, + \, 7 \, {\cal R} \, D^3 {\cal R}
\, + \,
9 \, (D {\cal R}) (\pa  {\cal R} ) ]
\ \ \ .
\nonumber
\end{eqnarray}

Methods for constructing the ${\cal L}_n$
and further properties of these operators
are given in reference \cite{cco} and in section eight below.
Here, we only note the following.
Under a conformal change
of $\bz$, the field $B$ transforms like a superaffine connection,
\begin{equation}
B_{\th^{\prime}} (\bz^{\prime} )
\ = \
{\rm e}^w \, \left[ B_{\th} (\bz ) + Dw  \right]
\ \ .
\end{equation}
Henceforth,
$[D^\prime-pB^\prime]C_p^\prime
= {\rm e}^{(p+1)w} [D -pB]C_p$ where $C_p$ has conformal weight ${p \over 2}$
and thereby we can
locally define a supercovariant derivative
\begin{eqnarray}
\NA_{(p)}\ :\ \ {\cal F}_{p} \! & \! &\longrightarrow\ \ {\cal F}_{p+1}
\nonumber
\\
C_{p}\!&\!&\longmapsto\NA_{(p)}C_p=[D-pB]C_p
\ \ .
\label{42}
\end{eqnarray}
Writing
$\NA_{(p)}^l \equiv \NA_{(p+l-1)} \cdot \cdot \cdot \NA_{(p+1)}
\NA_{(p)}$, the factorisation equation (\ref{45})
for the super Bol operators reads
\begin{equation}
\label{43}
{\cal L}_n \ = \ \NA_{(-n)}^{2n+1}
\ \ .
\end{equation}

It should be emphasized that superprojective structures
exist on compact SRS's of any genus and that superprojective connections
are globally defined on such spaces \cite{cco}.
Thereby, the lemma and corollary
stated above also hold there. By contrast, the covariant derivative
defined by eq.(\ref{42}) only exists locally. In fact, the quantity $B$ is not
invariant under superprojective changes of the coordinate
$\bZ$; under a conformal change
of $\bz$, the field $B$ transforms like a superaffine connection
(which quantities only exist globally on SRS's of genus one).
Nevertheless, the local definition (\ref{42}) is useful at intermediate
stages \cite{dfiz} and we will use it again in the next section.

\section{Ex.2: Operators parametrized by conformal fields}

Consider a fixed $n \in {\bf N}$ and
some tensors
$W_k\in{\cal F}_k$ (with $1\leq k\leq 2n+1$).
In analogy to the expression (\ref{43}), we can locally introduce
covariant operators $M_{W_k} ^{(n)}$ of order $2n+1-k$ in terms
of the fields $W_k$ and the
covariant derivative (\ref{42}):
we define them by
\begin{eqnarray}
M^{(n)}_{W_k}\ : \ {\cal F}_{-n}
\! & \! & \longrightarrow\ {\cal F}_{n+1}
\nonumber
\\
C_{-n} \! &\!&\longmapsto \
\sum_{l=0}^{2n+1-k}\beta_{kl}^{(n)}\left(\NA_{(k)}^l W_k\right)
\NA_{(-n)}^{2n+1-k-l}C_{-n}
\ \ ,
\label{199}
\end{eqnarray}
where the
$\beta_{kl}^{(n)}$ denote some complex numbers.
These operators are linear in $W_k$ and its derivatives up to order
$2n+1-k$.
Under a superconformal change of coordinates they transform as
$M^{(n)}_{W_k}\to {\rm e}^{(n+1)w} \, M^{(n)}_{W_k} \, {\rm e}^{nw}$.

\begin{prop}
There exist numerical coefficients
$\beta^{(n)}_{kl}$, normalized to $\beta^{(n)}_{k0}=1$, such that
the operators $M^{(n)}_{W_k}$ depend on $B$ only through the
superprojective connection
${\cal R}=\partial B-BDB$.
The coefficients
are explicitly given by
the following expressions where we distinguish between $k,l$ even
and odd:
\begin{equation}
\label{46}
\begin{array}{rclrcl}
\beta_{2k,2l}^{(n)}&=&{{\ds{n-k\choose l}{k+l-1\choose l}}\over
{\ds{2k+l-1\choose l}}} \quad, &
\beta_{2k,2l+1}^{(n)}&=&{{\ds{n-k\choose l}{k+l\choose l+1}}\over
{\ds{2k+l\choose l+1}}}
\quad ( 1 \leq k \leq n )
\\
\beta_{2k+1,2l}^{(n)}&=&{{\ds{n-k\choose l}{k+l\choose l}}\over
{\ds{2k+l\choose l}}} \quad , &
\beta_{2k+1,2l+1}^{(n)}&=&{{\ds{n-k\choose l+1}{k+l\choose l}}\over
{\ds{2k+l+1\choose l+1}}}
\quad ( 0 \leq k \leq n )
\ .
\end{array}
\end{equation}
\end{prop}

\noindent
The proof is by construction and uses a variational argument
\cite{ws} \cite{dfiz}:
we vary $B$
subject to the condition $\delta{\cal R}=0$ and require
$\delta M^{(n)} _{W_k} = 0$. The condition $\delta{\cal R}=0$
implies
\begin{equation}
\label{38}
(\partial \delta B)=(\delta B)(DB)+B(D\delta B)
\ \ .
\end{equation}
To proceed further, we need the following results.
{}From eq.(\ref{38}) and the definition of the covariant derivative,
one can derive the following operatorial relations:
\begin{eqnarray}
\NA_{(p)}(\delta B) & = & - (\delta B) \NA_{(p-1)}+(\NA_{(1)} \delta B)
\nonumber
\\
\NA_{(p)}(\NA_{(1)} \delta B) & = & (\NA_{(1)}\delta B)\NA_{(p-2)}
\label{41}
\\
\NA_{(p+1)} \NA _{(p)} (\delta B ) & = & (\delta B)
\NA _{(p)} \NA_{(p-1)}
\ \ .
\nonumber
\end{eqnarray}
Using these relations,
one shows that
\begin{equation}
\label{40}
\de\NA_{(p)}^l=d_l^{(p)}(\delta B)\NA_{(p)}^{l-1}
+b_l^{(p)}(\NA_{(1)}\delta B)\NA_{(p)}^{l-2}
\end{equation}
with coefficients
\begin{equation}
\label{39}
\begin{array}{rclrcl}
d_{2n}^{(p)} & = & -n \qquad  \quad \,\ , & \qquad
b_{2n}^{(p)}&=&-n(p+n-1)\quad,\\
d_{2n+1}^{(p)}&=&-(p+n)\quad , & \qquad
b_{2n+1}^{(p)}&=&-n(p+n)\quad.
\end{array}
\end{equation}
Indeed, one has $\delta\NA_{(p)}=-p\,\delta B$
(i.e. $d_1^{(p)}=-p,\,b_1^{(p)}=0$)
and eq.(\ref{40}) is then proved by induction: one is led to the
recursion relations $d_{l+1}^{(p)}=-(d_l^{(p)}+p+l)$
and $b_{l+1}^{(p)}= d_l^{(p)}+b_l^{(p)}$ with solutions
given by eqs.(\ref{39}).

Now, we are ready to prove the proposition. Variation of the
operator $M^{(n)}_{W_k}$ leads to the recursion relations
\begin{eqnarray}
(a)\qquad\beta_{k,l+1}^{(n)}d_{l+1}^{(k)}+(-)^{k+l}
\beta_{k,l}^{(n)}d_{2n+1-k-l}^{(-n)}&=&0 \qquad \ \ {\rm for} \ \quad
0\leq l\leq 2n-k
\nonumber\\
\,(b)\qquad\beta_{k,l+2}^{(n)}b_{l+2}^{(k)}+\phantom{(-)^{k+l}}
\beta_{k,l}^{(n)}b_{2n+1-k-l}^{(-n)}&=&0\qquad \ \ {\rm for} \ \quad
0\leq l\leq 2n-k-1
\nonumber
\end{eqnarray}
with $\beta_{k,0}^{(n)}=1$.  One first shows that $(b)$ follows from
$(a)$ by using the explicit form of the coefficients $d_l^{(p)}$
and $b_l^{(p)}$ given by eqs.(\ref{39}). Then, one solves $(a)$ and
finds the solution (\ref{46}).
\hfill      $\Box$

\begin{cor}
Let ${\cal W}_k$
(with $1\leq k\leq 2n+1$) be a quasi-primary field of weight ${k \over 2}$.
Then, the operator
$M^{(n)}_{{\cal W}_k}$, as defined by its action on a quasi-primary field
${\cal C}_{-n}$,
\begin{equation}
M_{{\cal W}_k}^{(n)} \,
{\cal C} _{-n}  \ \equiv \
\sum_{l=0} ^{2n+1-k} \,
\beta_{kl} ^{(n)} \, \left(
D_{\TH} ^{l} {\cal W}_k \right)
D_{\TH} ^{2n+1-k-l}
{\cal C}_{-n}
\end{equation}
with
$\beta_{kl} ^{(n)}$ given by eq.(\ref{46}),
transforms linearly under a superprojective change of coordinates:
\begin{equation}
\left( \,
M_{{\cal W}_k}^{(n)}\,
{\cal C}_{-n}\,\right)^{\prime}\ = \
(D_\Theta\Theta^\prime)^{-(n+1)}
\left(\,
M_{{\cal W}_k}^{(n)}\,
{\cal C}_{-n}\,\right)
\ \ .
\end{equation}
\end{cor}
This operator is related to the operator
$M^{(n)}_{W_k}$
by a change of variables in analogy to eq.(\ref{44}):
\[
M_{{\cal W}_k}^{(n)}\,
{\cal C}_{-n} \ = \
(D \Theta )^{-(n+1)}
M_{W_k}^{(n)}\,
C_{-n}
\quad \quad \quad
{\rm with}
\quad \quad
{\cal W}_k(\bZ) \, \equiv \, (D\Theta)^{-k} \, W_k (\bz )
\ \ .
\]

As an illustration of our results,
we give explicit expressions for the simplest
$M_{W_k}^{(n)}$:
\begin{eqnarray}
M_{W_{2n+1}}^{(n)} & = & W_{2n+1} \qquad , \qquad
M_{W_{2n}}^{(n)} \ = \ W_{2n} D + {1 \over 2} \, (DW_{2n} )
\nonumber \\
M_{W_{2n-1}}^{(n)} & = & W_{2n-1} \pa + {1 \over 2n-1} \, (DW_{2n-1}) D
 + {n \over 2n-1} \, (\pa W_{2n-1})
\\
M_{W_{2n-2}}^{(n)} & = & W_{2n-2} D^3 + {1 \over 2}  (DW_{2n-2}) \pa
 + {1 \over 2}  (\pa W_{2n-2}) D
 + {n \over 2(2n-1)}  (D^3 W_{2n-2})
 + {n^2 \over 2n-1}   W_{2n-2} {\cal R}
{}.
\nonumber
\end{eqnarray}

\section{Classification of covariant operators}

By adding the covariant operators considered in the last two sections,
we obtain the operator (\ref{gen}) which is again covariant and
of order $2n+1$. In fact, the procedure leading to the expression (\ref{gen})
can be reversed to show that
all normalized linear differential operators of order $2n+1$ which
are superholomorphic and superconformally covariant
can be cast into the form (\ref{gen}) with $W_1 = 0 = W_3$.
In particular, this means that all normalized and conformally
covariant operators come from M\"obius covariant ones.

\begin{theo}[Classification theorem]
Let (\ref{50})
be the local form of a
normalized differential operator of order $2n+1$.
According to lemma 3.1,
there exist transformation laws for
the coefficients $a_k$
(under superconformal changes of coordinates) such that
${\cal L}^{(n)}:{\cal F}_{-n}\to{\cal F}_{n+1}$.
If the $a_k$ are chosen to transform in this way,
one can find a reparametrization of these coefficients in terms of
superconformal fields
$W_k \in {\cal F}_k$
(with $1 \leq k \leq 2n+1$)
such that ${\cal L}^{(n)}$ is given by
the expression (\ref{gen}).

The $W_k$ are polynomials in the $a_k$ and their derivatives.
These relations are invertible and allow one to express
the $a_k$ as differential polynomials
in ${\cal R}$ and in the $W_k$
(with coefficients that are differential polynomials in ${\cal R}$).
\end{theo}

To summarize, any holomorphic and covariant operator of the form
(\ref{50}) can be parametrized by a superprojective connection ${\cal R}$
and $2n+1$ superconformal fields $W_1, ... , W_{2n+1}$ (which are
differential polynomials in the $a_k$).

The proof is by construction. In fact,
with eq.(\ref{49a}),
\begin{equation}
a_3\ = \ \frac{1}{2} \, n(n+1)\, \CR +\frac{1}{2} \,  Da_2
\ \ ,
\end{equation}
the $W_k$ are easily found in terms of the $a_k$ by setting to zero
the coefficient of $D^{2n+1-k}$ in ${\cal L}-{\cal L}^{(n)}$.
For the first seven fields, we find
\begin{eqnarray}
W_1 & = & 0
\nonumber \\
W_2 & = & a_2
\nonumber \\
W_3 & = & 0
\nonumber \\
W_4 & = & a_4-{1\over 6}n(n^2-1)D\CR- \frac{1}{2} (n-1)\partial a_2
\label{899}  \\
W_5 & = & a_5-\frac{1}{2} Da_4-{1\over 12}(n-1)D^3 a_2
          -{1\over 4}n(n^2-1)\partial \CR -{1\over 6}(3n+2)\CR a_2
\nonumber \\
W_6 & = & a_6-{1\over 5}(n-2)Da_5-{2\over 5}(n-2)\partial a_4
          +{1\over 10}(n-1)(n-2)\partial^2 a_2
          -{1\over 60}(n-2)(11n-1)\CR Da_2
\nonumber \\
    &   & \quad-{1\over 30}(n-2)
          (5n^2-3n-7)(D\CR)a_2+{1\over 20}n(n^2-1)(n-2)
          D^3\CR
\nonumber \\
W_7 & = & a_7-\frac{1}{2} Da_6-\frac{1}{2} (n-2)\partial a_5
          +{1\over 5}(n-2)D^3 a_4
          -{1\over 40}(n-1)(n-2)D^5 a_2
\nonumber \\
    &   & \quad-{1\over 10}(5n+3)(n-2)\CR a_4-{1\over 12}(n-2)^2(3n+2)
          (\partial\CR)a_2+{1\over 60}(n-2)(2n-11)\CR\partial a_2
\nonumber \\
    &   & \qquad+{1\over 20}n(n^2-1)(n-2)\partial^2\CR
          -{3\over 40}n(n^2-1)(n-2)\CR D\CR
\ \ .
\nonumber
\end{eqnarray}
These fields transform covariantly.

\section{Covariant bilinear operators}

The expression
$M_{W_k}^{(n)} C_{-n}$ can be viewed as the result of a bilinear and
covariant map $J$, i.e.
$M_{W_k}^{(n)} C_{-n} \propto J(W_k , C_{-n})$.
The mapping $J$ represents the graded extension of
Gordan's transvectant \cite{bol} \cite{gp} \cite{dfiz}.
It will reappear in the next sections in the context of the matrix
representation for linear covariant operators and in the Poisson
brackets of super $W$ algebras.

To define this extension, we proceed as for the definition of
$M_{W_k}^{(n)}$. First, we note that for any $\mu, \nu, m \in
{\bf Z}$, the map
\begin{eqnarray}
J^m _{\mu \nu}\ : \ {\cal F}_{\mu} \otimes {\cal F}_{\nu}
\! & \! & \longrightarrow\ {\cal F}_{\mu + \nu + m}
\nonumber
\\
(F, G) \! &\!&\longmapsto \
\sum_{l=0}^{m} \gamma_l^m (\mu , \nu ) \,
\left(\NA_{(\mu)}^{m-l} F\right)  \left(
\NA_{(\nu)}^{l} G \right)
\ \ ,
\label{700}
\end{eqnarray}
(with
$\gamma_l^m (\mu , \nu )$
denoting a numerical factor) is bilinear and covariant.
We then have the

\begin{prop}
There exist numerical coefficients
$\gamma_l^m (\mu , \nu )$
such that the operator $J_{\mu \nu} ^m$
depends on $B$ only through the
superprojective connection
${\cal R}=\partial B-BDB$.
These coefficients
are explicitly given by
\begin{equation}
\label{701}
\begin{array}{rclcrcl}
\gamma_{2l} ^{2m} (\mu , \nu ) & = &   (-1)^l
{\ds{m \choose l}} {\ds{1 \over (\mu)_{m-l} (\nu )_l }}
 & ,   &
\gamma_{2l+1} ^{2m} (\mu , \nu ) & = &   (-1)^{l + \mu}
{\ds{m-1 \choose l}} {\ds{m \over  (\mu)_{m-l} (\nu )_{l+1} }}
\\
& & & & & &
\\
\gamma_{2l} ^{2m+1} (\mu , \nu ) & = &   (-1)^l
{\ds{m \choose l}} {\ds{1 \over (\mu)_{m+1-l} (\nu )_l  }}
 & ,  &
\gamma_{2l+1} ^{2m+1} (\mu , \nu ) & = &  - (-1)^{l + \mu}
{\ds{m \choose l}} {\ds{1 \over (\mu)_{m-l} (\nu )_{l+1} }}
,
\end{array}
\end{equation}
where we made use of the Pochammer symbol
\[
(r)_0 \equiv 1 \qquad , \qquad (r)_l = r(r+1)\cdot \cdot \cdot
(r+l-1)
\ \ .
\]
\end{prop}

\noindent
The proof
proceeds along the lines of the proof of proposition 5.1:
the coefficients are determined by requiring that
$\delta J_{\mu \nu}^m (F,G) =0$ for variations
$\delta B$ satisfying
$\delta{\cal R}=0$.
{}From eq.(\ref{40}), we get the following recursion relations
for the coefficients $\gamma \equiv \gamma (\mu , \nu)$:
\begin{eqnarray*}
\gamma_{2l} ^{2m}  & = & -(-1)^{\mu} \;
{m+\mu-l \over l} \
\gamma_{2l-1} ^{2m}
\ \ \ , \ \ \
\gamma_{2l+1} ^{2m}  \ = \ (-1)^{\mu} \ {m-l \over l+\nu}  \;
\gamma_{2l} ^{2m}
\\
\gamma_{2l} ^{2m+1}  & = & (-1)^{\mu} \ {m-l+1 \over l} \;
\gamma_{2l-1} ^{2m+1}
\ \ \ \ , \ \ \
\gamma_{2l+1} ^{2m+1} \ = \ -(-1)^{\mu} \ {m+\mu-l \over \nu+l} \;
\gamma_{2l} ^{2m+1}
\  .
\end{eqnarray*}
Using a convenient choice for $\gamma_0$, one is led to the solution
(\ref{701}).
\hfill       $\Box$

The last proposition can be reformulated by saying that
\begin{equation}
J^m _{\mu \nu} ({\cal F} , {\cal G} ) \ \equiv \
\sum_{l=0}^{m} \gamma_l^m (\mu , \nu ) \,
\left(D_{\TH} ^{m-l} {\cal F} \right)  \left(
D_{\TH} ^l {\cal G} \right)
\label{702}
\end{equation}
(with $\gamma$ given by eqs.(\ref{701})) is a quasi-primary superfield
of weight ${1 \over 2} (\mu + \nu +m)$ if
${\cal F}$ and
${\cal G}$ are quasi-primary of weight ${\mu \over 2}$
and ${\nu \over 2}$, respectively.
In other words, the operator (\ref{702})
is bicovariant with respect to super M\"obius transformations.

The super Gordan transvectant
(i.e. the mapping $J_{\mu \nu}^m$ with $\gamma$ given by eqs.(\ref{701}))
has the symmetry properties
\begin{eqnarray}
J^{2m} _{\mu \nu} ( F , G ) & = & \ \ \, (-1)^{m + \mu \nu} \ \,
J^{2m} _{\nu \mu} ( G , F )
\\
J^{2m+1} _{\mu \nu} ( F , G ) & = &  - (-1)^{m + \mu \nu} \
J^{2m+1} _{\nu \mu} ( G , F )
\ \ .
\nonumber
\end{eqnarray}

As a simple example,
we consider $\mu = 2 = \nu, m=3$ and $F=G \equiv V \in {\cal F}_2$
for which case we get
\begin{eqnarray}
6 \, J^3 _{2 ,2} (V, V) & = &
2 \, V (\NA_{(2)} ^3 V) - 3 \,
(\NA_{(2)} V)
(\NA_{(2)} ^2 V)
\nonumber
\\
& = &
2 \, V (D^3  V) - 3\,
(D V) (\pa V) - 4 \, {\cal R} V^2
\ \ .
\label{703}
\end{eqnarray}

As noted at the beginning of this section, the quantity
$M_{W_k} ^{(n)} C_{-n}$ is a special case of $J^m _{\mu \nu} (F,G)$:
in fact, for
\[
\mu =  k
\quad  , \quad
\nu  =  -n
\quad , \quad
m  = 2n+1-k
\quad , \quad
F  =  W_{k}
\quad , \quad
G  =  C_{-n}
\]
we find
\begin{equation}
\label{704}
J^m _{\mu \nu} (  F ,  G ) \ = \
\left\{
\begin{array}{rll}
{\ds{(k-1)! \over n!}} &
M_{W_{2k}} ^{(n)} C_{-n}
& \quad  {\rm for} \quad \mu = 2k
\\
&  &
\\
{\ds{k! \over n!}} &
M_{W_{2k+1}} ^{(n)} C_{-n}
& \quad  {\rm for} \quad \mu = 2k+1
\end{array}
\right.
\end{equation}

\section{Matrix representation}

The differential equation
$F_{2n+1} \, = \, {\cal L}^{(n)} f_{2n+1}$
which is of order $2n+1$
is equivalent to a system of $2n+1$ first-order equations.
The latter system can be cast into matrix form which
provides an elegant and efficient way
for determining explicit expressions
for the covariant operators. Moreover, the matrix representation
exhibits most clearly
the underlying algebraic structure which is
due to the covariance with respect to
super M\"obius transformations.

Let us first discuss the case of the super Bol operator ${\cal L}_n$.
We consider it in its factorized form (see eq.(\ref{43})),
\begin{equation}
\label{801}
{\cal L}_n \ = \ (D-nB) \cdot \cdot \cdot (D +nB)
\ \ ,
\end{equation}
where $B$ represents a superaffine connection.
For $n=0,1,...$, the scalar equation
\begin{equation}
\label{802}
F_{2n+1} \ = \ {\cal L}_n f_{2n+1}
\qquad {\rm with} \qquad
f_{2n+1} \in {\cal F}_{-n}
\end{equation}
is equivalent to the $(2n+1)\times (2n+1)$ matrix equation
\begin{equation}
\label{803}
\vec{F} \ = \ \tilde{Q} _n \vec{f}
\ \ ,
\end{equation}
with $\vec{F} \, = \,
\left( F_{2n+1} ,0,...,0 \right)^t$ and
$\vec{f} \, = \,
\left( f_1 ,...,f_{2n+1} \right)^t$.
Here,
\begin{equation}
\label{804}
\tilde{Q} _n  \ =\ -J_{-} + D {\bf 1} - B H
\ \ ,
\end{equation}
where ${\bf 1}$ stands for the unit matrix and
\begin{equation}
\label{805}
J_- \ = \
\left(
\begin{array}{cccccc}
0   & \cdots & \cdots  & \cdots   &   0  \\
1   & \ddots &         &          & \vdots   \\
0   & 1      & \ddots  &          & \vdots \\
\vdots &     & \ddots  & \ddots   &  \vdots  \\
0   &  \cdots   &   0   & 1   & 0
\end{array}
\right)
\ \ \ \ \ \ , \ \ \ \ \ \ \
H \ = \
\left(
\begin{array}{ccccc}
n   & 0     & \cdots  & \cdots & 0          \\
0   & n-1    & \ddots  &        & \vdots \\
\vdots   &       & \ddots  & \ddots   & \vdots  \\
\vdots   &       &         & \ddots & 0  \\
0   &  \cdots & \cdots &  0     & -n
\end{array}
\right)
 \ \ .
\end{equation}
The matrices
$J_-$ and $H$ satisfy
\begin{equation}
\label{807}
[ H , J_- ] = - J_-
\ \ .
\end{equation}
Together with the upper triangular matrix
\begin{equation}
\label{808}
J_+  \ = \
\left(
\begin{array}{ccccccc}
0      & n      &   0  & \cdots & \cdots & \cdots  & 0      \\
\vdots & \ddots &   -1 &        &        &         & \vdots  \\
\vdots &        &\ddots& n-1    &         &        & \vdots  \\
\vdots &        &      & \ddots &  -2     &        &  \vdots \\
\vdots &        &      &        & \ddots  & \ddots &   0     \\
\vdots &        &      &        &         & \ddots & -n       \\
0      & \cdots &\cdots& \cdots & \cdots  & \cdots & 0
\end{array}
\right) \
\ \ ,
\end{equation}
which satisfies
\begin{equation}
\label{809}
[ H , J_+ ] =  J_+
\qquad , \qquad  \{ J_+ , J_- \} = H
\ \ ,
\end{equation}
they generate an $osp(1|2)$ algebra. In fact \cite{drs}, these
matrices represent the superprincipal embedding
$osp(1|2)_{{\rm ppal}}$
of $osp(1|2)$ into the superalgebra $sl(n+1|n)$.
For an elaboration on the algebraic structure, we refer
to the appendix.
The matrix $\tilde{Q}_n$
can be cast into
a canonical form $Q_n$ by conjugation with a group
element $N\in OSP(1|2)_{{\rm ppal}}
\subset SL(n+1|n)$. This makes the
dependence on the superprojective structure manifest:

\begin{theo}[Matrix representation for the super Bol operator]
The scalar operator ${\cal L}_n$ is equivalent to the matrix operator
$Q_n$ defined by
\begin{equation}
\label{810}
Q_n \ \equiv \
\hat{N}^{-1} \tilde{Q} _n  N \ =\ -J_{-} +D {\bf 1}
- {\cal R} J_+ ^2
\ \ .
\end{equation}
Here, ${\cal R} = \pa B - B(DB)$ represents
a superprojective connection and
\begin{equation}
\label{811}
N \ = \ {\rm exp} \left\{ -B J_+ - (DB) J_+ ^2 \right\}
\qquad , \qquad
\hat{N}  \ =\
{\rm exp} \left\{ +B J_+ - (DB) J_+ ^2 \right\}
\ \ .
\end{equation}
\end{theo}

\noindent
As for the proof, we first note that
$J_+$ ($J_+^2$) is odd (even) with respect to
the ${\bf Z}_2$-grading $i+j \, ({\rm mod} \, 2)$ of matrix elements
defined in the appendix.
In eq.(\ref{811}), these generators are multiplied with odd (even)
parameters and therefore the corresponding expressions represent
well defined elements of the supergroup $SL(n+1 |n)$.

The element $\hat{N} \equiv {\rm exp} \{ \hat{M} \}$ follows
from $N \equiv {\rm exp} \{ M \}$
by changing the sign of the anticommuting part of the algebra
element $M$.
(This operation represents
an automorphism of the superalgebra \cite{drs}.)
The consideration of $\hat{N}$ is necessary in the conjugation
(\ref{810}), because the operator $\tilde{Q}_n$
has a grading different
from the one considered for the superalgebra
(e.g. $\tilde{Q}_n$ contains odd elements on the diagonal).

The result (\ref{810})
is a simple consequence of the relations
\begin{eqnarray*}
\hat{N}^{-1} J_-  N & = & J_{-} - BH - (DB) J_+ + 2B(DB) J_+^2
\\
\hat{N}^{-1} D  N & = & D {\bf 1}
- (DB) J_+ + \left[ B(DB) -(\pa B) \right] J_+^2
\\
\hat{N}^{-1} BH   N & = & B  \left[ H - BJ_+ - 2(DB) J_+^2  \right]
\ \ ,
\end{eqnarray*}
which follow from eqs.(\ref{807}) and (\ref{809}).

The matrix (\ref{810}) still describes the operator ${\cal L}_n$
\cite{ds}. In fact,
one can transform an equation of the form
(\ref{803}) by upper triangular matrices
\[
N( \bz ) \ = \
\left(
\begin{array}{ccccc}
1   & \ast  & \ast & \cdots & \ast  \\
0   & 1      & \ast  & \cdots   &   \ast  \\
\vdots   &       & \ddots   &\ddots   & \vdots  \\
\vdots   &       &          & \ddots  & \ast    \\
0   &  \cdots &\cdots & 0   & 1
\end{array}
\right)
\ \ \ ,
\]
which leads to the equation
\begin{equation}
\label{806}
\vec{F}^{\prime} \ = \
\tilde{Q} _n ^{\prime}
\vec{f}^{\prime}
\ \ ,
\end{equation}
with
$\vec{F}^{\prime} \ \equiv \ \hat{N} \, \vec{F}\, , \,
\vec{f} ^{\prime} \ \equiv \ N \, \vec{f}$ and
$\tilde{Q}_n ^{\prime} \ \equiv \ \hat{N} \, \tilde{Q}_n \, N^{-1}$.
Since
$F_{2n+1} ^{\prime} \ =\  F_{2n+1}$ and
$f_{2n+1} ^{\prime} \ = \  f_{2n+1}$, the matrix equation (\ref{806}) is
equivalent to the scalar equation (\ref{802}), i.e.
$\tilde{Q}_n ^{\prime}$ still
describes the operator ${\cal L}_n$.
\hfill   $\Box$

To conclude our discussion of the super Bol operator,
we remark that
the matrix representation (\ref{810})
for ${\cal L}_n$
is equivalent to a representation in terms of
an $(n+1) \times (n+1)$ matrix involving $D{\cal R}$
and the operators $D, \pa $ as given in
reference \cite{cco}.

Next, we
consider the most general covariant operators ${\cal L}^{(n)}$
as discussed in section 3 and 6.

\begin{theo}[Matrix representation for covariant operators]
For $n=0,1,...$,
the most general normalized and
covariant operator of order $2n+1$
is given by the scalar equation
\begin{equation}
\label{830}
F_{2n+1} \ = \ {\cal L}^{(n)}  \, f_{2n+1}
\qquad {\rm with} \qquad
f_{2n+1} \in {\cal F}_{-n}
\ \ ,
\end{equation}
which is obtained after elimination of $f_1,...,f_{2n}$ from the
matrix equation
\begin{equation}
\label{831}
\left( F_{2n+1} ,0,...,0 \right) ^t
\ = \ Q ^{(n)}
\left( f_1 ,...,f_{2n+1} \right) ^t
\ \ .
\end{equation}
Here, $Q^{(n)}$ is the
$(2n+1) \times (2n+1)$ matrix
\begin{equation}
\label{832}
Q ^{(n)}  \ =\ -J_{-} + D {\bf 1} + \sum_{k=1}^{2n+1} V_{k+1} M_k
\ \ ,
\end{equation}
where
$V_3 \equiv {\cal R} \, , \,
V_k \in {\cal F}_k$ for $k=2$ and for $4\leq k \leq 2n+1$ and
$M_k=(M_1)^k$ with
\begin{equation}
\label{833}
M_1\ = \
\left(
\begin{array}{ccccccc}
0      & n      &   0  & \cdots & \cdots & \cdots  & 0      \\
\vdots & \ddots &    1 &        &        &         & \vdots  \\
\vdots &        &\ddots& n-1    &         &        & \vdots  \\
\vdots &        &      & \ddots &   2     &        &  \vdots \\
\vdots &        &      &        & \ddots  & \ddots &   0     \\
\vdots &        &      &        &         & \ddots & n       \\
0      & \cdots &\cdots& \cdots & \cdots  & \cdots & 0
\end{array}
\right) \
\ \ .
\end{equation}
\end{theo}

\noindent
A few comments concerning this result and its interpretation are in order.
First, it should be noted that
the representation (\ref{810}) for the super Bol operator
is a special case of the representation (\ref{832}),
since $J_+^2 = - (M_1)^2$.

The representation (\ref{832}) obtained here as a generalization
of (\ref{810}) is analogous to results
based on the work of Drinfel'd and Sokolov
and on constrained Wess-Zumino-Witten models (see \cite{bal}
\cite{dfiz} \cite{bx}
\cite{drs} and
references therein).
The superconformal
fields $V_k$ are introduced in the matrix representation
along with the highest weight generators of the
$osp(1|2)_{{\rm ppal}}$ subalgebra of $sl(n+1 |n)$.
These are the generators $M_p$ which satisfy the graded commutation relation
\[
\left[ \left. J_+ , M_p \right\} \right. \ = \ 0
\ \ ,
\]
where $M_p$ is characterized by its $H$-eigenvalue,
\[
\left[ \left. H , M_p \right\} \right. \ = \ p M_p
\ \ .
\]
The smallest value is $p=1$  and
one easily finds for $M_1$ the matrix (\ref{833})
as a solution
of the previous equations.
The integer powers of $M_1$ still belong to the superalgebra
and one readily shows that
\begin{equation}
\left[ \left. H , (M_1)^p \right\} \right. \ = \ p (M_1)^p
\qquad , \qquad
\left[ \left. J_+ , (M_1)^p \right\} \right. \ = \ 0
\ \ ,
\end{equation}
from which equations we conclude that $M_p=(M_1)^p$.

In conclusion, we note that the
superconformal fields $V_k$ occurring in eq.(\ref{832})
represent a parametrization of the operator ${\cal L}^{(n)}$
which is equivalent to the one in terms of the $W_k$ discussed
in section 6:
the two sets of superfields are related to each other by differential
polynomials. For concreteness, we illustrate the situation for $n=3$.
In this case, eq.(\ref{832}) yields
\[
{\cal L}^{(3)} \ = \
{\cal L}_3 +
M_{W_2}^{(3)} + ... +
M_{W_7}^{(3)}
\ \ ,
\]
where ${\cal L}_3$ is the super Bol operator and
$M_{W_k}^{(3)}$ the operators of section 5 with $W_k$ depending on
the $V_k$ and their derivatives according to
\begin{eqnarray}
W_2 & = & 12 V_2
\nonumber    \\
W_3 & = & 0
\nonumber    \\
W_4 & = & 10 V_4  + 44 (V_2)^2
\nonumber    \\
W_5 & = & 5  V_5
 \\
W_6 & = & 2  V_6  + 48 (V_2)^3 + 36 V_2 V_4
\nonumber    \\
W_7 & = &  V_7  + 18 V_2 V_5 + \frac{9}{5} \left[
2 \, V_2 ( D^3  V_2) - 3\,
(D V_2) (\pa V_2) - 4 \, {\cal R} (V_2)^2  \right]
\ \ .
\nonumber
\end{eqnarray}
Up to an overall factor,
the expression in brackets coincides with
the super Gordan transvectant $J_{2,2}^3$ applied to the pair of fields
$(V_2, V_2)$, see eq.(\ref{703}).
Obviously,
$V_k \in {\cal F}_k$ implies
$W_k \in {\cal F}_k$ and the last set of relations is invertible
since the leading term of $W_k$ is always given by $V_k$.

For later reference, we also summarize the results for
$n=1$ and $n=2$. In these cases, one finds, respectively,
${\cal L}^{(1)}  =  {\cal L}_1 +  M_{W_2}^{(1)}$ with
\begin{equation}
\label{836}
W_2 \ = \ 2 V_2
\end{equation}
and
${\cal L}^{(2)} \ = \ {\cal L}_2 + M_{W_2}^{(2)} + ... + M_{W_5}^{(2)}$ with
\begin{equation}
\label{835}
W_2  =   6 V_2
\quad  , \quad
W_3   =   0
\quad  , \quad
W_4   =   4 V_4  + 8 (V_2)^2
\quad  , \quad
W_5   =   4  V_5
\ \ .
\end{equation}
For any value of $n$, we have
$W_2 = n(n+1) V_2$ where the coefficient $n(n+1)$ represents the sum
of elements of the matrix $M_1$.

\section{Classical super $W$ algebras}

Classical super $W_n$ algebras represent non-linear extensions of the
classical super Virasoro algebra. We recall that the latter is generated
by the super stress tensor ${\cal T}$ which transforms
like a superprojective connection
${\cal R}$ or rather like the combination (\ref{49a}),
\begin{equation}
\label{900}
\tilde{a}_3 ^{(n)} \ \equiv \ a_3 ^{(n)} \, - {1\over 2} \,  Da_2 ^{(n)}
\ = \
\iota_n \, {\cal R}
\qquad \quad {\rm with} \quad
\iota_n \, \equiv \,
\frac{1}{2}  \, n (n+1)
\ \ ,
\end{equation}
which expression involves the coefficients
$a_2 ^{(n)}$ and $a_3 ^{(n)}$ of a normalized
covariant operator of order $2n+1$.
The Poisson bracket defining the super Virasoro algebra is then given by
\cite{pm}
\begin{equation}
\label{901}
\{ \tilde{a}_3^{(n)} (\bz^{\prime}) , \tilde{a}_3 ^{(n)} (\bz ) \}
\ = \ {1 \over 2} \;
\left[ \, \iota_n
D^5  + 3 \tilde{a}_3^{(n)} \pa + (D \tilde{a}_3^{(n)}) D
+ 2 ( \pa  \tilde{a}_3^{(n)})
\, \right]
\de (\bz - \bz^{\prime} )
\ \ ,
\end{equation}
where $\de (\bz - \bz^{\prime} ) = (\th - \th^{\prime} )
\, \de (z - z^{\prime} )$.
By substituting
$\tilde{a}_3^{(n)} = \iota_n {\cal R}$
into this relation, we see that the operator on the r.h.s. coincides with
${1\over 2} \iota_n {\cal L}_2$ where ${\cal L}_2$ is the super Bol operator
(\ref{898}).

Arguing along the lines of the bosonic theory
\cite{pan} \cite{dfiz} \cite{bx}, one can say that the super $W_n$ algebra
is generated by the super stress tensor
$\tilde{a}_3^{(n)}$  and the superconformal fields $W_k$ (with
$2 \leq k \leq 2n+1$) which parametrize covariant operators of order $2n+1$
according to section six.
It follows from the results of Gel'fand and Dickey and their
generalization to odd superdifferential
operators \cite{mr} that the Poisson brackets
between these generators form a closed algebra.
For the $n=1$ and $n=2$ cases, these brackets have recently been
constructed \cite{hn}
(see also \cite{ik} \cite{v}).
Starting from normalized
covariant operators ${\cal L}^{(n)}$, the authors
of references \cite{hn} and \cite{ik} found that these brackets
take the simplest form if they are written in terms
of superconformal fields
$V_k$ which are specific differential polynomials
in the coefficients $a_k ^{(n)}$ of
${\cal L}^{(n)}$. For the examples studied,
these combinations are precisely the combinations $V_k$ that
we encountered for the matrix representation of covariant operators
in section 8. Thus, these fields seem to be better suited
for the parametrization of super $W$ algebras than the combinations $W_k$
discussed in section 6 for the classification
of covariant operators. In any case,
our results provide a general method for determining these combinations.

Our results admit a further application to the formulation of super
$W$ algebras. The $n=1$ and $n=2$
Poisson brackets derived in reference \cite{hn}
by virtue of a long and tedious calculation
involve a large number of terms which depend on the generators
$\tilde{a}_3^{(n)}$ and $V_k$. All of these contributions
can be rewritten in a compact way in terms of the covariant operators
constructed in the present work (i.e. ${\cal L}_n,M^{(n)} _{W_k},
J_{\mu \nu}^m$) and of some covariant trilinear operators.
In the following, we briefly summarize
the results of reference \cite{hn} while
emphasizing and elucidating the underlying algebraic structure.

For $n=1$, one starts from the covariant operator
\[
{\cal L}^{(1)} \ = \ D^3 + a_2^{(1)} D + a_3^{(1)}
\]
and eqs.(\ref{900})(\ref{836})(\ref{899}) then yield
\begin{eqnarray}
{\cal R} & = &
\tilde{a}_3 ^{(1)} \ = \ a_3 ^{(1)} \, - {1\over 2} \, Da_2 ^{(1)}
\nonumber  \\
2 V_2 & = & W_2 \ = \ a_2 ^{(1)}
\ \ .
\label{902}
\end{eqnarray}
The super stress tensor
${\cal T} \equiv \tilde{a}_3 ^{(1)}$ and the superconformal field
${\cal J} \equiv 2 V_2 $
satisfy the algebra\footnote{In the $\{ {\cal T} , {\cal J} \}$-bracket
of reference \cite{hn}, there is an obvious sign
error which we have corrected here (cf. eq.(\ref{78}) below).}
\begin{eqnarray}
\{ {\cal J} (\bz^{\prime}) , {\cal J} (\bz ) \}  & = &
2 \, \left[ \, D^3  +  {\cal T} \, \right]
\de (\bz - \bz^{\prime} )
\nonumber  \\
\{ {\cal T} (\bz^{\prime}) , {\cal T} (\bz ) \}  & = &
{1 \over 2} \, \left[ \,
D^5  + 3 {\cal T} \pa + (D {\cal T} ) D
+ 2 ( \pa  {\cal T} )
\, \right]
\de (\bz - \bz^{\prime} )
\label{903}
\\
\{ {\cal T} (\bz^{\prime}) , {\cal J} (\bz ) \}  & = &
-  \, \left[ \,
(\pa {\cal J} ) +
{\cal J}  \pa - {1\over 2} (D {\cal J} ) D
\, \right]
\de (\bz - \bz^{\prime} )
\ \ .
\nonumber
\end{eqnarray}
Noting that ${\cal T} = {\cal R}$ in the present case,
we recognize the super Bol operators ${\cal L}_1$ and ${\cal L}_2$
on the r.h.s. of the first and second equations,
respectively. The operator in the last relation is covariant,
since it can be obtained from the super Gordan transvectant
$J_{\mu \nu}^m $ as a linear operator
$-2J_{2, -2}^2( {\cal J} , \cdot )$.
This relation represents the transformation law of the superconformal
field ${\cal J}$ under a superconformal change of coordinates
generated by the stress tensor ${\cal T}$.
For a field $C_k \in {\cal F}_k$, it generalizes to
\begin{equation}
\label{895}
\{ {\cal T} (\bz^{\prime}) , C_k (\bz ) \}  \ = \
\lambda_k \,
J_{k, -2}^2( C_k  , \cdot )  \,
\de (\bz - \bz^{\prime} )
\qquad \quad {\rm with} \quad
\lambda_k = - (-1)^k \, k
\ \ .
\end{equation}

Component field expressions for $n = 1$
follow from the $\th$-expansions
\begin{equation}
\label{894}
{\cal J} (\bz ) = J(z) + \th G^2 (z)
\qquad , \qquad
{\cal T} (\bz ) = G^1(z) + \th T (z)
\ \ ,
\end{equation}
where $T(z)$ represents the ordinary stress tensor.
These component fields together with their commutation relations following
from eqs.(\ref{903})
define a representation
of the $N=2$ superconformal algebra.

For $n=2$, one considers
\[
{\cal L}^{(2)} \ = \ D^5 + a_2^{(2)} D^3 + a_3^{(2)} D^2 +
a_4^{(2)} D + a_5^{(2)}
\]
and eqs.(\ref{900})(\ref{835})(\ref{899}) now lead to
\begin{eqnarray}
{\cal T} & \equiv &
3 {\cal R} \ = \ \tilde{a}_3 \ = \ a_3 - {1\over 2}  Da_2
\nonumber  \\
{\cal J} & \equiv &
6 V_2 \ = \ W_2 \ = \ a_2
\label{904} \\
{\cal U} & \equiv &
4 V_4 \ = \ W_4 - 8 (V_2)^2
\ = \ a_4 - {1\over 3}  Da_3
- {1\over 3}  \pa a_2  - {2\over 9} (a_2)^2
\nonumber  \\
{\cal W} & \equiv &
4 V_5 \ = \ W_5
\ = \ a_5 - {1\over 2}  Da_4
- {1\over 2}  \pa a_3  + {1\over 6}  D^3 a_2
- {4\over 9}  a_2 \tilde{a}_3
\ \ ,
\nonumber
\end{eqnarray}
where $a_k \equiv a_k ^{(2)}$.
(In order to avoid a confusion with the Schwarzian derivative, the
superfield $S$ of reference \cite{hn} is being
referred to as ${\cal U}$.)
The $n=2$ super Poisson brackets read
\begin{eqnarray}
\{ {\cal J} (\bz^{\prime}) , {\cal J} (\bz ) \}  & = &
6 \; {\cal L}_1 \;
\de (\bz - \bz^{\prime} )
\qquad \qquad , \qquad \qquad
\{ {\cal T} (\bz^{\prime}) , {\cal T} (\bz ) \}  \ = \
{3 \over 2} \; {\cal L}_2 \;
\de (\bz - \bz^{\prime} )
\label{906}
\\
&&
\nonumber  \\
\{ {\cal T} (\bz^{\prime}) , {\cal J} (\bz ) \}  & = &
-2 \;
J_{2, -2}^2( {\cal J} , \cdot )  \;
\de (\bz - \bz^{\prime} )
\qquad , \qquad
\{ {\cal T} (\bz^{\prime}) , {\cal U} (\bz ) \}  \ = \
- 4 \;
J_{4, -2}^2( {\cal U} , \cdot )  \;
\de (\bz - \bz^{\prime} )
\nonumber  \\
\{ {\cal T} (\bz^{\prime}) , {\cal W} (\bz ) \}  & = &
\ 5 \;
J_{5, -2}^2( {\cal W} , \cdot )  \;
\de (\bz - \bz^{\prime} )
\qquad , \qquad
\{ {\cal J} (\bz^{\prime}) , {\cal W} (\bz ) \}  \ = \
\ 2 \;
J_{4, -1}^2( {\cal U} , \cdot )  \;
\de (\bz - \bz^{\prime} )
\nonumber  \\
&&
\nonumber  \\
\{ {\cal U} (\bz^{\prime}) , {\cal U} (\bz ) \}  & = &
- \left[  {2 \over 3}
{\cal L}_3  +
{1 \over 27}  M^{(3)} _{w_4}   -
{1 \over 90}  M^{(3)} _{w_7}  \right]
\de (\bz - \bz^{\prime} )
\quad , \quad
\{ {\cal U} (\bz^{\prime}) , {\cal J} (\bz ) \}  \; = \;
2  {\cal W}
\de (\bz - \bz^{\prime} )
\nonumber  \\
&&
\nonumber  \\
\{ {\cal W} (\bz^{\prime}) , {\cal U} (\bz ) \}  & = &
\left[ {16 \over 3}
J_{2,-4}^6 ( {\cal J} ,  \cdot )  +  20
J_{5,-4}^3 ( {\cal W} ,  \cdot )  -
J_{6,-4}^2 ( w_6 , \cdot )
 +  {5 \over 27}
K_{2,4,-4}^2  ({\cal J} , {\cal U} , \cdot )  \right]
\de (\bz - \bz^{\prime} )
\nonumber  \\
\{ {\cal W} (\bz^{\prime}) , {\cal W} (\bz ) \}  & = &
-  \left[ \,
{1\over 6} \, {\cal L}_4 \, + \, {2 \over 9} \,
J_{4,-4}^5 ( w_4 ,  \cdot )
\, + \, {26 \over 9} \,
J_{7,-4}^2 ( {\cal J}{\cal W} ,  \cdot )
\right.
\nonumber  \\
& & \left. \qquad  \qquad  \quad \quad
- \, {5 \over 126} \,
K_{2,5,-4}^2  ({\cal J} , {\cal W} , \cdot )
\, + \, {1 \over 540} \,
K_{2,2,-4}^5  ({\cal J} , {\cal J} , \cdot ) \, \right]
\, \de (\bz - \bz^{\prime} )
\ ,
\nonumber
\end{eqnarray}
with
\begin{equation}
w_4 \ = \ 90 \, {\cal U} - 2 \, {\cal J}^2
\quad  ,  \quad
w_6 \ = \ {8 \over 81} \, {\cal J}^3 - {28 \over 9} \, {\cal J} {\cal U}
 \quad  ,  \quad
w_7 \ = \ 20 \, {\cal J}{\cal W} + 27 \, J_{2,2}^3 ( {\cal J} , {\cal J} )
\label{907}
\end{equation}
and
\begin{eqnarray}
K_{2,4,-4}^2  ( {\cal J} , {\cal U} , \cdot ) & = &
{\cal J} ( \NA  {\cal U}) \NA  -2\,
( \NA {\cal J}) {\cal U} \NA  +2\,
( \NA^2 {\cal J}) {\cal U}
- \,
{\cal J} ( \NA^2 {\cal U})  - {3 \over 2} \,
( \NA {\cal J}) (\NA {\cal U} )
\label{908}
\\
K_{2,5,-4}^2  ({\cal J} , {\cal W} , \cdot  ) & = & 4
{\cal J} ( \NA {\cal W}) \NA  -10 \,
( \NA {\cal J}) {\cal W} \NA  -5\,
( \NA^2 {\cal J}) {\cal W}
\, +2\,
{\cal J} ( \NA^2 {\cal W})  + 7 \,
( \NA {\cal J}) (\NA {\cal W} )
\nonumber  \\
K_{2,2,-4}^5  ( {\cal J} , {\cal J} , \cdot ) & = &
47  \, {\cal J} (\NA ^3 {\cal J} ) \NA^2 \, - \,
{141 \over 2} \, (\NA {\cal J} )
(\NA ^2 {\cal J} ) \NA^2 \, + \,
{\cal J} (\NA ^4 {\cal J} ) \NA
\, - \, {3 \over 2} \,
(\NA ^2 {\cal J} )^2 \NA
\nonumber  \\
& &
+ \, {45 \over 2} \,
(\NA {\cal J} ) (\NA ^3 {\cal J} ) \NA \, + \, 24
{\cal J}  (\NA ^5 {\cal J} )
\, - \, 2
(\NA ^2 {\cal J} ) (\NA ^3 {\cal J} ) \, - \, 46
(\NA {\cal J} ) (\NA ^4 {\cal J} )
\ .
\nonumber
\end{eqnarray}
In eqs.(\ref{906}),
the commutator of ${\cal T}$ with itself and the one of ${\cal T}$
with the other fields represent
special cases of the relations (\ref{901}) and  (\ref{895}),
respectively.
The operators (\ref{908}) are examples of trilinear and
covariant
operators,
\begin{eqnarray}
K_{\mu \nu \rho}^m \ : \
{\cal F}_{\mu} \otimes
{\cal F}_{\nu} \otimes
{\cal F}_{\rho}
\! & \! & \ \longrightarrow \
{\cal F}_{\mu + \nu + \rho +m}
\nonumber  \\
(F,G,H)
\! & \! & \ \longmapsto \
K_{\mu \nu \rho}^m (F,G,H)
\ \ ,
\end{eqnarray}
which only depend on superconformal fields and
on the superprojective connection ${\cal R}$
(i.e. on the superprojective structure).
Whereas the bilinear operators discussed in section seven
are unique (up to an overall normalization) this is in general
not the case for multilinear generalizations.
For any value of $n$, the Poisson brackets are at most cubic
in the generators $W_k$ ($2 \leq k \leq 2n+1$) \cite{dfiz} and therefore
they involve at most quadrilinear covariant operators.

\section{Projection to component fields}

All superfields $F(\bz )$ admit a $\th$-expansion
$F(\bz ) = f (z) + \th \, \psi (z)$ where the component field
$f$ has the same
Grassmann parity as $F$. In particular, the superprojective
connection $\CR$ can be written as \cite{cco}
\begin{equation}
\label{200}
\CR _{z\th} (\bz ) \ = \ \frac{i}{2} \,
\rho_{z\th} (z) \, + \, \th \, [\frac{1}{2} r_{zz} (z) ]
\ \ ,
\end{equation}
where $r_{zz}$ corresponds to an ordinary projective connection on the Riemann
surface underlying the SRS.

In order to project down the supercovariant derivative (\ref{42})
to component field expressions, we note that the superconformal transformation
$\bZ = \bZ (\bz)$ relating the generic coordinates $\bz$ to the
projective coordinates $\bZ$ satisfies $DZ = \TH D\TH$.
This equation implies $(D\TH)^2 = \pa Z + \TH \pa \TH$ and therefore
\begin{equation}
\label{198}
B \ \equiv \ D \, {\rm ln} \, D\TH \ = \ {1 \over 2} \, D \, {\rm ln} \, \pa Z
\, + \, {1 \over 2} \, D \biggl[ \frac{\TH \pa \TH}{\pa Z} \biggr]
\ \ .
\end{equation}
In the following, we
denote the lowest component of a generic superfield $F$ by $F \! \! \mid$.
Writing $B = {i \over 2} \beta + \th [{1 \over 2} b]$ and
using eq.(\ref{198}), we get
\begin{equation}
\label{195}
b \ = \
\pa \, {\rm ln} \, \pa Z \! \! \mid
\ + \ {\rm SUSY}
\ \ ,
\end{equation}
where `SUSY' stands for the contributions which are due to supersymmetry.
By virtue of eq.(\ref{42}) we obtain
the component field expression
\begin{eqnarray}
\left( \nabla_{(p)} ^2 C_p \right) \! \! \mid
& = &
[ \, \pa - {p \over 2} \, b \, ] \; C_p \! \! \mid
\ - \ {i \over 2} \, \beta \, (DC_p) \! \! \mid
\nonumber \\
& = & \CD _{({p \over 2})} \, c_{{p\over 2}} + \ {\rm SUSY}
\ \ .
\label{201}
\end{eqnarray}
Here, $C_p \! \! \mid \equiv c_{\frac{p}{2}}$
represents an ordinary conformal field of weight ${p\over 2}$ and
$\CD _{({p \over 2})}
\equiv \pa - {p \over 2} \, (\pa \, {\rm ln} \, \pa
Z \! \! \mid )$ denotes the ordinary covariant derivative
acting on such fields (see e.g. \cite{dfiz}).
{}From the $\th$-expansions of $\CR$ and $B$ and from
eqs.(\ref{197}),(\ref{195}),
we conclude that
$b$ and $\beta$ give rise to the following local form of
the projective connection $r$:
\begin{equation}
\label{196}
r \ = \ \pa^2 \, {\rm ln} \, \pa Z \! \! \mid \, - \, {1 \over 2} \,
(\pa \, {\rm ln} \, \pa Z \! \! \mid )^2 \ + \ {\rm SUSY}
\ \ .
\end{equation}

The projection of the super Bol operator ${\cal L}_n$ has been discussed
in reference \cite{cco} and it was shown that
\begin{equation}
\label{202}
\left( D {\cal L} _{n-1} C_{-(n-1)} \right) \! \! \mid \ = \
L_n \, c_{-\frac{n-1}{2}} \ + \ {\rm SUSY}
\ \ ,
\end{equation}
where $L_n$ denotes the ordinary Bol operator of
order $n$. For instance, for $n=3$,
\[
\left( D {\cal L} _{2} C_{-2} \right) \! \! \mid \ = \
L_3 \, c_{-1} \ + \ {\rm SUSY}
\qquad {\rm with} \qquad
L_3 = \pa^3 + 2 r \pa + (\pa r )
\ \ .
\]

For the operator $M_{W_k}^{(n)}$ introduced in eq.(\ref{199}),
we can extract the purely bosonic
contribution in the same way:
\begin{equation}
\label{203}
\left( D M_{W_{2k}}^{(n-1)} C_{-(n-1)} \right) \! \! \mid \ = \
M_{w_{k}}^{(n)} c_{-\frac{n-1}{2}}  \ + \ {\rm SUSY}
\ \ .
\end{equation}
Here, $W_{2k} \! \! \, \mid \equiv w_k$ is an ordinary conformal field
of weight $k$ and
$M_{w_{k}}^{(n)}$ represents the known bosonic result \cite{dfiz} \cite{gp},
\begin{equation}
\label{204}
M_{w_{k}}^{(n)} \ = \ \sum_{l=0}^{n-k}
\alpha_{kl}^{(n)}
\left( \CD_{(k)} ^l w_k \right) \, \CD_{(-{n-1 \over 2})} ^{n-k-l}
\qquad {\rm with } \qquad
\alpha_{kl}^{(n)} \ = \ \frac{{k+l-1 \choose l}
{n-k \choose l}}{
{2k+l-1 \choose l}}
\end{equation}
and with $\CD$ denoting the covariant derivative
introduced in eq.(\ref{201}). We note that the derivation of eq.(\ref{203})
makes use of the relation
\begin{equation}
\label{205}
\beta_{2k,2l}^{(n-1)} \, + \,
\beta_{2k,2l-1}^{(n-1)} \ = \
\alpha_{kl}^{(n)}
\qquad \qquad
( \ 0 \leq l \leq n-k-1 \ )
\ \ .
\end{equation}

Proceeding in the same way for the bilinear operator $J_{2\mu , 2\nu} ^m$,
we find
\begin{equation}
\label{206}
DJ_{2\mu , 2 \nu}^{2m-1} (F_{2\mu} , G_{2\nu} ) \! \! \mid
\ = \ j_{\mu \nu}^m (f_{\mu} , g_{\nu} )\ + \ {\rm SUSY}
\ \ .
\end{equation}
Here,
$F_{2\mu} \! \! \mid \equiv f_{\mu} \, , \,
G_{2\nu} \! \! \mid \equiv g_{\nu}$
are ordinary conformal fields of weight $\mu$ and $\nu$, respectively,
and $j$ denotes Gordan's transvectant \cite{bol} \cite{gp} \cite{dfiz},
\begin{equation}
\label{207}
j_{\mu \nu}^m (f_{\mu} , g_{\nu})
= \sum_{l=0}^{m}
\de_{l}^m (\mu , \nu ) \, \left( {\cal D}_{(\mu)} ^{m-l} f_{\mu} \right)
\left( {\cal D}_{(\nu)} ^l g_{\nu} \right)
\quad {\rm with } \quad
\de_{l}^m (\mu , \nu )  =
(-1)^l \,
{m \choose l} \, {1 \over (\mu)_{m-l} (\nu)_l}
{}.
\end{equation}
The result (\ref{206}) follows from the relation
\begin{equation}
\label{208}
\gamma_{2l} ^{2m-1} (2\mu , 2\nu) +
\gamma_{2l-1} ^{2m-1} (2\mu , 2\nu) \ = \
\de_{l}^m (\mu , \nu )
\qquad \quad (\ 1 \leq l \leq m-1 \ )
\ \ .
\end{equation}
By construction, the operators (\ref{202}),(\ref{203}) and (\ref{206})
depend on
the quantity
$\pa \, {\rm ln} \, \pa Z \! \! \mid$
only through the projective connection (\ref{196}).

Component field expressions for
the super Poisson brackets of section nine immediately follow from the
$\th$-expansions.
Here, we only note that
the transformation law of superconformal fields
$C_k (\bz) = c_{{k\over 2}} (z) + \th c_{\frac{k+1}{2}} (z)$
as described by the super bracket (\ref{895})
implies the usual transformation law of ordinary conformal fields under
conformal changes of coordinates $z \to z^{\prime} (z)$.
In fact, from eqs.(\ref{895})
and (\ref{894}), it follows that
\begin{eqnarray}
\label{78}
\{ T(z^{\prime}) , c_k (z) \} & =&
\left[ \, (\pa c_k) + k \, c_k \pa \, \right] \, \de (z -z^{\prime} )
\\
& = & k \, j_{k,-1} ^1 (c_k , \cdot ) \, \de (z - z^{\prime} )
\ \ ,
\nonumber
\end{eqnarray}
where $j$ denotes Gordan's transvectant (\ref{207}).

So far we have outlined the
projection of our superfield results and we have verified
that they encompass the known bosonic expressions. For the presentation
of complete component field results, it may be convenient to interpret
a superfield $F(\bz ) = f(z) + \th \psi (z)$ as a doublet
$\left( f(z) ,\psi (z) \right) ^t$ and to describe the action of a
superdifferential operator ${\cal L}$ on it
by a $2 \times 2$ matrix whose elements
are ordinary differential operators. This approach was considered in reference
\cite{v} where some basic aspects of covariance and super $W$ algebras have
been
discussed in terms of this formalism.

\section{Concluding remarks}
We have discussed superconformally covariant differential operators
and their relevance for classical super $W$ algebras. Our discussion
was restricted to operators of odd degree in $D$, i.e.
${\cal L}=D^{2n+1}+\dots$. Let us now comment on operators
of even degree, i.e. operators of the form ${\cal L}=D^{2n}+
a_1 D^{2n-1}+a_2 D^{2n-2}+\dots =\partial^n+a_1 D^{2n-1}
+a_2\partial^{n-1}+\dots$.
One first notices that for these operators it is {\em not} possible
to eliminate the coefficient $a_1$ of the subleading term
in the way it was done for bosonic operators or for the
supersymmetric operators discussed in this paper.
One can however eliminate $a_2$ and if one requires
$a_2$ to stay zero under superconformal changes of coordinates,
one finds that ${\cal L}$ has to map
${\cal F}_{1-n}$ to ${\cal F}_{1+n}$. Yet, if one tries
to generalize the super Bol operators and considers
$\nabla^{2n}_{(p)}$, one finds that this expression
depends on the connection
$B$ not only through the superprojective connection
${\cal R}$ for any choice of $p$. This means in particular that
the super Bol lemma does not hold in this case
and that the analysis
presented in this paper does not carry over directly to
normalized superdifferential operators of even degree.
They do not seem to be relevant for classical super $W$ algebras
(see also \cite{v} \cite{fr}), but, interestingly enough, the even operator
${\cal L}^{(n)} D$ (where ${\cal L}^{(n)} = D^{2n+1} +...$
is a normalized covariant
operator) represents the Lax operator for the generalized $N=2$ super
KdV hierarchy \cite{ik}.

Our discussion of differential operators and classical $W$ algebras
has been based on $N=1$ superconformal fields.
The super $W$ algebras considered in section nine are however
$N=2$ algebras. For instance, in the $n=1$ case, one has the
$N=2$ super Virasoro algebra involving the super stress tensor
and the superfield $a_2$ which contains
the $U(1)$ current $J$ and the second supercharge $G^2$ as component
fields. As mentioned
in section three, we can consistently set $a_2$ to zero which
reduces the symmetry algebra to $N=1$. Thus, the $N=1$ formalism used here
encompasses both the $N=1$ and $N=2$ cases.
To exhibit the $N=2$ structure more clearly for $a_2 \neq 0$,
one could use an $N=2$ superfield formalism \cite{wip}.
In that case, the $osp(1|2)$ algebra should be
replaced by an $sl(2|1)$ algebra according to the discussion
of the spin content of super $W$ algebras in
reference \cite{drs}.

As we were finishing this work, we became aware of reference \cite{bsa}
where a general formula for the basic singular vectors of the
Neveu-Schwarz (NS) algebra was presented. If written in terms of matrices,
this formula uses the same representation
of $osp(1 | 2)$ than the one we encountered in section eight.
In terms of our notation in section 8,
the main result of reference \cite{bsa} reads
\begin{equation}
\vec{F} \ = \  \left[ \, - J_{-} \, + \, \sum_{p=0}^{n}  \, {2p \choose p} \,
G_{-{2p+1 \over 2}} \, \left( {t \over 2} \right)^p  J_+^{2p} \, \right] \,
\vec{f}
\ \ .
\end{equation}
Here, $f_1, ..., f_{2n+1}$ and $F_{2n+1}$ belong to the Verma module
$V_{(c,h)}^{{\rm NS}}$ of the NS algebra; more specifically,
$f_{2n+1} \equiv | h \! \! >$ represents a highest weight vector and
$F_{2n+1} \equiv | \psi_{1,q} \! \! >$ a singular vector
of fundamental type (i.e. all the others can be obtained from these ones
\cite{bsa}). $G_r$ denotes the fermionic generator of the NS algebra
and $t$ a complex parameter.

In conclusion, we mention that some conformally covariant operators
on ordinary Riemann surfaces (like Bol operators)
admit a natural generalization to higher
dimensional Riemannian manifolds \cite{erb}. The corresponding
covariant operators can be constructed by virtue of a matrix representation
or by using homomorphisms of Verma modules. Superconformally covariant
operators as discussed in \cite{cco} and in the present work
should allow for an analogous generalization to higher dimensional
supermanifolds and our results should provide the appropriate basis for
their construction.

\vskip 2.0truecm

{\bf \Large Acknowledgements}

\vspace{5mm}

It is a great pleasure to thank Fran\c cois Delduc
for his precious advice on graded algebras and for his
valuable comments on the matrix representation.

\newpage

\appendix
\section{Some facts about superalgebras}

In this appendix, we collect
some general results on superalgebras
relevant for our discussion of the algebraic structure of
matrix operators in section 8.

By definition, the superalgebra
$sl(n+1 | n)$ is the graded algebra of $(2n+1) \times (2n+1)$
matrices $M$ with vanishing supertrace:
\begin{equation}
\label{a1}
{\rm str}_1 \, M \ \equiv \ \sum_{i=1}^{n+1} M_{ii} -
\sum_{i=n+2}^{2n+1} M_{ii} \ = \ 0
\ \ .
\end{equation}
The matrix $M$ is made up of blocks,
\begin{equation}
\label{a2}
M \ = \
\left(
\begin{array}{cc}
A   & C \\
D   & B
\end{array}
\right)
\ \ ,
\end{equation}
with $A,B$ even and $C,D$ odd.

In section 8, we encountered another representation of $sl(n+1|n)$
which is based on a different definition of the grading and trace.
In this representation, one
associates a ${\bf Z}_2$-grading $i+j \ ({\rm mod} \ 2)$
to the element $M^{\prime}_{ij}$ of the matrix $M^{\prime} \in
sl(n+1 | n)$.
Then, there are no more even and odd blocks as in eq.(\ref{a2}), rather
the elements of a row or column of $M^{\prime}$ are alternatively odd and even.
The supertrace is now given by an alternating sum over the diagonal elements of
the matrix:
\begin{equation}
\label{a3}
{\rm str}_2\,M^{\prime}\ \equiv\ \sum_{i=1}^{2n+1}\,(-1)^{i+1}\,
M^{\prime}_{ii}
\ \ .
\end{equation}
Note that in both expressions (\ref{a1}) and (\ref{a3}) the number
of plus (minus) signs is $n+1$ ($n$).
The graded commutator is defined by
\begin{equation}
\label{a4}
\left[ \left. M^{\prime} , N^{\prime} \right\} _{ik} \right.
\ \equiv \
\sum_{j=1}^{2n+1} \left( M^{\prime}_{ij} N^{\prime}_{jk}
\, - \, (-1)^{(i+j)(j+k)} \,
N^{\prime}_{ij} M^{\prime}_{jk}\right)
\end{equation}
and one has
\[
{\rm str}_2 \,
\left[ \left. M^{\prime} , N^{\prime} \right\} \right.  \ = \ 0
\ \ .
\]
The two different representations of
$sl(n+1 | n)$ are related by a similarity transformation,
\begin{equation}
\label{a5}
M^{\prime} \ = \
G^{-1} M G
\ \ ,
\end{equation}
where $G$ is a permutation matrix,
$G_{ij} = \de_{i P(j)}$ with
\begin{eqnarray*}
P(2i+1) & = & i+1 \qquad \quad \ \ \,
{\rm for} \quad 0 \leq i \leq n
\\
P(2i) & = & n+i+1 \qquad {\rm for} \quad 1 \leq i \leq n
\ \ .
\end{eqnarray*}

In the following, we consider supertraceless matrices
of the type $M^{\prime}$.
To simplify the notation, we will suppress the
`prime'.

Let $E_{ij}$ denote the matrix
\begin{equation}
\label{a6}
\left( E_{ij} \right) _{kl} \ = \
\de_{ik} \de_{jl}
\ \ .
\end{equation}
As a basis of the Cartan subalgebra of $sl(n+1 | n)$,
we can take the matrices
\[
h_i \ = \ E_{ii} + E_{i+1,i+1}
\qquad {\rm for} \qquad
1 \leq i \leq 2n
\ \ ,
\]
while
the simple roots can be chosen to be fermionic and represented by
\[
e_i \ = \ E_{i, i+1} \quad ({\rm positive \ \, roots})
\qquad , \qquad
f_i \ = \ E_{i+1, i} \quad ({\rm negative \ \, roots})
\ \ .
\]
Then, the superprincipal embedding of the graded algebra
$osp(1|2)$ in
$sl(n+1 | n)$ is given by
$J_- \equiv \sum_{i=1}^{n} f_i$ and $H \, , \, J_+$ (as
given by eqs.(\ref{805}),(\ref{808})) supplemented
by the bosonic generators
$X_{\pm} = J_{\pm}^2$.
These matrices satisfy the graded commutation relations (\ref{807})
and (\ref{809}) from which it follows that
\begin{equation}
[ H, X_{\pm} ] = \pm 2X_{\pm}
\qquad , \qquad
[ X_+, X_- ] = - H
\qquad , \qquad
\lbrack J_\mp, X_\pm\rbrack =\pm J_\pm
\ \ .
\end{equation}


\newpage

\begin{thebibliography}{22}
\newcommand{\artref}[4]{{\sc #1}, {\it #2} {\bf #3} #4}
\newcommand{\bookref}[2]{{\sc #1}, #2}


\bibitem{bol}
\bookref
{P.Gordan}{``Invariantentheorie", (Teubner, Leipzig, 1887) ;}

\bookref
{E.J.Wilczynski}{``Projective differential geometry of curves
and ruled surfaces", (Chelsea Publ., New York, 1905) ;}

\artref
{G.Bol}{Abh.Math.Sem.Hamburger Univ.}{16}{(1949) 1 ;}

\artref
{C.Teleman}{Comm.Math.Helv.}{33}{(1959) 206 .}

\bibitem{gp}
\artref
{S.Janson and J.Peetre}{Math.Nachr.}{132}{(1987) 313 ;}

\artref
{B.Gustafsson and J.Peetre}{Nagoya Math.J.}{116}{(1989) 63 ;}

\bookref
{B.Gustafsson and J.Peetre}{``M\"obius invariant differential
operators on Riemann surfaces", in ``Function spaces, operators and
non-linear analysis", L.P\"aiv\"arinta, ed., Pitman Research Notes
in Math. Series (Longman, Harlow 1989) .}

\bibitem{cco}
\bookref
{F.Gieres}{``Conformally covariant operators on Riemann surfaces
(with applications to conformal and integrable models)",
preprint CERN-TH.5985/91, MPI-Ph/91-39,
to appear in Int.J.Mod.Phys.A .}

\bibitem{pan}
\artref
{P.Mathieu}{Phys.Lett.}{B208}{(1988) 101 ;}

\artref
{I.Bakas}{Commun.Math.Phys.}{123}{(1989) 627 ;}

\artref
{R.Viswanathan}{Phys.Lett.}{B244}{(1990) 419 ;}

\artref
{Q.Wang, P.K.Panigrahi, U.Sukhatme and W.-Y.Keung}{Nucl.Phys.}{B344}{
(1990) 196 .}

\bibitem{dfiz}
\artref
{P.Di Francesco, C.Itzykson and J.-B.Zuber}{Commun.Math.Phys.}{140}{
(1991) 543;}

\artref
{M.Bauer, P.Di Francesco, C.Itzykson and J.-B.Zuber}{Nucl.Phys.}{B362}{
(1991) 515 .}

\bibitem{bx}
\artref
{L.Bonora and C.S.Xiong}{Int.J.Mod.Phys.}{A7}{(1992) 1507 .}

\bibitem{df}
\bookref
{D.Friedan}{in ``Unified String Theories", Santa Barbara Workshop,
M.B.Green and D.Gross, eds. (World Scientific, 1986) ;}

\artref
{A.M.Baranov, Yu.I.Manin, I.V.Frolov and A.S.Schwarz}{Commun.Math.Phys.}
{111}{(1987) 373 ;}

\artref
{L.Crane and J.M.Rabin}{Commun.Math.Phys.}{113}{(1988) 601 ;}

\artref
{M.Batchelor and P.Bryant}{Commun.Math.Phys.}{114}{(1988) 243 .}

\bibitem{ws}
\artref
{W.Scherer}{Lett.Math.Phys.}{17}{(1989) 45 .}

\bibitem{drs}
\artref
{F.Delduc, E.Ragoucy and P.Sorba}{Commun.Math.Phys.}{146}{
(1992) 403.}

\bibitem{ds}
\artref
{V.G.Drinfel'd and V.V.Sokolov}{Journ.Sov.Math.}{30}{(1985) 1975 .}

\bibitem{bal}
\artref
{J.Balog, L.Feher, P.Forgacs, L.O'Raifeartaigh and A.Wipf}{Ann.Phys.}{203}{
(1990) 76 .}

\bibitem{pm}
\artref
{P.Mathieu}{J.Math.Phys.}{29}{(1988) 2499 .}

\bibitem{mr}
\artref
{Yu.I.Manin and A.O.Radul}{Commun.Math.Phys.}{98}{(1985) 65 .}

\bibitem{hn}
\artref
{K.Huitu and D.Nemeschansky}{Mod.Phys.Lett.}{A6}{(1991) 3179 .}

\bibitem{ik}
\bookref
{T.Inami and H.Kanno}{``$N=2$ super $W$ algebras and generalized
$N=2$ super KdV hierarchies based on Lie superalgebras",
preprint YITP-K-928 (May 1991) .}

\bibitem{v}
\artref
{R.R.Viswanathan}{Int.J.Mod.Phys.}{A7}{(1992) 4501 .}

\bibitem{fr}
\artref
{J.M.Figueroa-O'Farrill and E.Ramos}{Commun.Math.Phys.}{145}{(1992) 43 .}

\bibitem{wip}
\bookref
{F.Gieres and S.Theisen}{work in progress .}

\bibitem{bsa}
\artref
{L.Benoit and Y.Saint-Aubin}{Lett.Math.Phys.}{23}{(1991) 117 .}

\bibitem{erb}
\artref
{M.G.Eastwood and J.W.Rice}{Commun.Math.Phys.}{109}{(1987) 207 ;}

\bookref
{R.J.Baston and M.G.Eastwood}{in ``Twistors in Mathematics and Physics",
T.N.Bailey and R.J.Baston, eds. (Cambridge University Press, 1990) .}

\end{thebibliography}
\end{document}